\documentclass{pasa}%

\usepackage{framed}
\usepackage[lined,ruled,commentsnumbered]{algorithm2e}

\usepackage{pdflscape}
\usepackage{array}

\usepackage{multirow}
\usepackage[normalem]{ulem}

\usepackage{enumerate}

\usepackage{verbatimbox}

\SetStartEndCondition{ }{}{}%
\SetKwProg{Def}{}{}{}%
\SetKwFunction{Range}{range}
\SetKw{KwTo}{in}\SetKwFor{For}{for}{\string:}{}%
\SetKwIF{If}{ElseIf}{Else}{if}{:}{elif}{else:}{}%
\SetKwFor{While}{while}{:}{fintq}%
\SetKwComment{tcc}{`{}`{}`{}}{'{}'{}'{}}
\AlgoDontDisplayBlockMarkers\SetAlgoNoEnd\SetAlgoNoLine%

\usepackage{tikz}
\usetikzlibrary{shapes.geometric, arrows, positioning, decorations.markings}

\title[Measuring in situ MWA beams]{In situ measurement of MWA primary beam variation using ORBCOMM}

\usepackage[authoryear]{natbib}
\bibpunct{(}{)}{;}{a}{}{,}
\def\Unimelb{$^{1}$}
\def\Curtin{$^{2}$}
\def\ASU{$^{3}$}
\def\USydney{$^{4}$}
\def\UToronto{$^{5}$}
\def\UWisc{$^{6}$}
\def\UW{$^{7}$}
\def\UWA{$^{8}$}
\def\CAASTRO{$^{9}$}

\author[J.~L.~B.~Line]{J.~L.~B.~Line\Unimelb$^,$\CAASTRO$^,$\thanks{{\href{malito:jack.line@curtin.edu.au}{jack.line@curtin.edu.au}}},
B.~McKinley\Curtin$^,$\CAASTRO, 
J.~Rasti\Unimelb$^,$\CAASTRO,
M.~Bhardwaj\Unimelb$^,$\CAASTRO,
R.~B.~Wayth\Curtin$^,$\CAASTRO, 
R.~L.~Webster\Unimelb$^,$\CAASTRO,
D.~Ung\Curtin,
D.~Emrich\Curtin,
L.~Horsley\Curtin,
A.~Beardsley\ASU,
B.~Crosse\Curtin,
T.~M.~O.~Franzen\Curtin,
B.~M.~Gaensler\USydney$^,$\CAASTRO$^,$\UToronto,
M.~Johnston-Hollitt\Curtin,
D.~L.~Kaplan\UWisc, 
D.~Kenney\Curtin,
M.~F.~Morales\UW, 
D.~Pallot\UWA,
K.~Steele\Curtin,
S.~J.~Tingay\Curtin$^,$\CAASTRO,
C.~M.~Trott\Curtin$^,$\CAASTRO,
M.~Walker\Curtin,
A.~Williams\Curtin, 
C.~Wu\UWA
\\
\\
\affil{\Unimelb School of Physics, The University of Melbourne, Melbourne, VIC 3010, Australia}
\affil{\Curtin International Centre for Radio Astronomy Research, Curtin University, Bentley, WA 6102, Australia}
\affil{\ASU School of Earth and Space Exploration, Arizona State University, Tempe, AZ 85287, USA}
\affil{\USydney Sydney Institute for Astronomy, School of Physics, The University of Sydney, NSW 2006, Australia}
\affil{\UToronto Dunlap Institute for Astronomy and Astrophysics, University of Toronto, ON, M5S 3H4, Canada}
\affil{\UWisc Department of Physics, University of Wisconsin--Milwaukee,
Milwaukee, WI 53201, USA}
\affil{\UW Department of Physics, University of Washington, Seattle, WA 98195, USA}
\affil{\UWA International Centre for Radio Astronomy Research, University of Western Australia, Crawley 6009, Australia}
\affil{\CAASTRO ARC Centre of Excellence for All-sky Astrophysics (CAASTRO)}
}

\jid{PASA}
\doi{10.1017/pas.\the\year.xxx}
\jyear{\the\year}

\usepackage{hyperref} 
\hypersetup{
  colorlinks   = true, 
  urlcolor     = blue, 
  linkcolor    = blue, 
  citecolor    = black 
}

\begin{document}

\begin{abstract}
We provide the first in situ measurements of antenna element (tile) beam shapes of the Murchison Widefield Array (MWA), a low radio-frequency interferometer and an SKA\thanks{\url{https://www.skatelescope.org/project/}} precursor. Most current MWA processing pipelines use an assumed beam shape, errors in which can cause absolute and relative flux density errors, as well as polarisation `leakage'. This makes understanding the primary beam of paramount importance, especially for sensitive experiments such as a measurement of the 21$\,$cm line from the epoch of reionisation (EoR). The calibration requirements for measuring the EoR 21$\,$cm line are so extreme that tile to tile beam variations may affect our ability to make a detection. Measuring the primary beam shape from visibilities alone is challenging, as multiple instrumental, atmospheric, and astrophysical factors contribute to uncertainties in the data. Building on the methods of~\citet{Neben2015}, we tap directly into the receiving elements of the MWA before any digitisation or correlation of the signal. Using ORBCOMM satellite passes we are able to produce all-sky maps for 4 separate tiles in the XX polarisation. We find good agreement with the cutting-edge `fully' embedded element (FEE) model of~\citet{Sokolowski2017}, and observe that the MWA beamformers consistently recreate beam shapes to within~$\sim1$dB in the reliable areas of our beam maps. We also clearly observe the effects of a missing dipole from a tile in one of our beam maps, and show that the FEE model is able to reproduce this modified beam shape. We end by motivating and outlining additional onsite experiments to further constrain the primary beam behaviour.

\end{abstract}


\maketitle


\section{Introduction}
\label{sec:intro}
The Murchison Widefield Array (MWA) is a low radio-frequency interferometer located in the western Australian outback, at the Murchison Radio-astronomy Observatory (MRO). One of the observational strengths of the MWA is its substantial field of view. Each receiving element (tile) in the interferometer consists of a $4\times4$ grid of bow-tie dipoles mounted on a $5\,\mathrm{m} \times 5\,\mathrm{m}$ reflective ground screen. Analogue beamformers are used to create and electronically steer the MWA primary beam, which has a full-width half-maximum of $\sim 25^\circ$ at 150$\,$MHz~\citep{Tingay2013}. The regular spacing of the dipoles means that the quantised beamformer delays are exactly correct for a set of pointings, reducing the complexity of the instrument. By using identical receiving elements, a number of computational simplifications can be made to calibration and imaging, as beam corrections can be made in image space, avoiding costly convolutions in visibility space.

Many calibration schemes make assumptions on beams/receiving elements~\citep[e.g.][]{Kazemi2013,Tasse2013}, and others explicitly use the beam shape during calibration/imaging~\citep[][specifically for the MWA]{Mitchell2008,Sullivan2012}. However, for these assumptions to be valid, each tile must actually produce identical beam shapes, and that beam shape must be correctly modelled.

The exact degree of precision required of a primary beam and model depends on the science case, but epoch of reionisation (EoR) science is perhaps in the greatest need of well understood beam shapes. Particularly, the spectral behaviour of the beam is a significant possible contaminant of a detection~\citep{Barry2016,Trott2016b}. \citet{Barry2016} state that the frequency response of an EoR experiment should have spectral features no larger than $10^{-5}$. This kind of spectral behaviour could be injected via calibration using an incorrect beam model.

Estimating the degree of precision of a beam model required for EoR science is an ongoing area of research. The effects of incorrect beam modelling on output data products can be subtle, and will change according to the calibration scheme used. To fully estimate the effects on an MWA EoR power spectrum experiment\citep[e.g.][]{Jacobs2016}, one has to take the following points into consideration:
\begin{itemize}
\item The primary beam is often used as a gridding kernel. An incorrect model would bias the gridded data used to either image or create a power spectrum
\item Errors on the beam model can be direction dependent. As the MWA primary beam is stationary during a 2 minute observation, primary calibrators will move through these errors during the observation. The MWA EoR observing strategy often employs the same pointing for $\sim30\,$minutes, potentially injecting time-dependent calibration errors
\item If the primary beam varies from tile to tile, the spatial scales that are measured by each baseline are affected differently. The effect of this upon a measured power spectrum is hard to estimate.
\end{itemize}

We plan to investigate the points above in future work, using simulated observations and power spectrum measurements to directly quantify the effects on potential EoR science.

As the accuracy of the beam model can affect science outputs, significant work has gone into both simulating and measuring the MWA primary beam. The MWA is a full Stokes instrument, and as such the beam model must accurately describe the instrument in all polarisations. Of particular interest is `leakage' from Stokes I to other polarisations, whereby flux density from Stokes I is transferred into other Stokes parameters. This effect is often elevation dependent, and can be attributed to an incorrect beam model~\citep{Lenc2016,Lenc2017}. 

The sophistication of the MWA's beam model has progressed through a number of stages. The first model was a simple analytical short-dipole radiation pattern multiplied by the tile array factor~\citep{Ord2010}. The tile array factor is a direction dependent function that describes the cumulative beam shape effects of the superposition of the individual dipoles, in this case generated purely through the geometrical $4\times4$ layout of the tile. The first advanced model was the `average' embedded element (AEE) model, presented by~\citet{Sutinjo2015a}. This model used numerical simulations generated using the commercially available FEKO\footnote{\url{https://www.feko.info/}} simulation package, including mutual coupling effects, to create an average dipole radiation pattern, which was assumed identical for all dipoles. In addition, the model includes mutual-coupling induced changes to dipole impedance, which affects the array factor depending on pointing direction.

Whilst this model was shown to reduce leakage,~\citet{Sutinjo2015a} showed that leakage still remained, and suggested that a `fully' embedded element (FEE) model was required. This was verified as the AEE model was used in calibration and image-based primary beam correction on GaLactic and Extragalactic All-sky MWA survey~\citep[GLEAM,][]{Wayth2015} data. Frequency/elevation dependent errors in the model produced polarisation leakage and incorrect absolute flux density of sources in the images~\citep[sometimes $>10$\%; see][for details]{Hurley-Walker2017}. Given catalogues such as GLEAM are often used to calibrate observations, a correct flux scale is of paramount importance. As a response, the FEE model was presented in~\citet{Sokolowski2017}. This model more rigorously takes into account mutual coupling, amongst other improvements. \citet{Sokolowski2017} were able to use GLEAM observations to show reduced leakage compared to the AEE model.

Using polarisation leakage to test beam performance is a valuable metric, but is a measure of average beam effects, rather than the beam pattern itself. To validate a beam model, one must directly measure the actual beam shape of individual tiles. The simplest way to do this is to have a radio emitter of known location and strength, and use the instrument to measure the flux density seen from the emitter from various locations. Recent advances have been made in using commercially available drones to fly a radio emitter to achieve this~\citep[e.g.][]{Ustuner2014,Chang2015,Picar2015,Pupillo2015,Paonessa2016}. These experiments have the advantage of being able to run at multiple frequencies and along a controlled flight path. The disadvantage is that these experiments happen in the near-field of the instrument; observations occur in the far-field. Even with this limitation, once the stability of the drones and repeatability of results improves, this approach could provide accurate beam measurements in the future.

Another option lies in using transiting satellites, bright in the MWA frequency band, as locatable emitters. \citet{Neben2015} have successfully used ORBCOMM satellites\footnote{\url{https://www.orbcomm.com/en/networks/satellite}} to measure an MWA tile beam shape in an experimental setup at the National Radio Astronomy Observatory in Green Bank, West Virginia. The constellation of ORBCOMM satellites operate at 137$\,$MHz, which does limit any experiment to a single frequency band, but the sky coverage and intrinsic brightness of the satellites allows a beam map to be made across the entire MWA primary beam. Satellites have the advantage of naturally being in the far-field of the instrument. \citet{Neben2016} were able to apply this method to the Hydrogen Epoch of Reionisation Array\footnote{HERA - \url{http://reionization.org/}}, to both measure beam shapes, and comment on the implications to their science goals. The success of Neben et al. has motivated a comparison of ORBCOMM and drone-based beam mapping techniques~\citep{Jacobs2017}, aimed towards being able to conduct reliable onsite experiments.

To date, no one has actually directly measured the individual MWA tile beam shapes onsite at the MRO. Given the discussion above, we suggest the ideal MWA beam measurement experiment should:
\begin{itemize}
\item occur onsite at the MRO using actual MWA tiles
\item cover the whole observable sky, in the far-field
\item measure multiple tiles simultaneously to establish similarities/differences
\item span the entire frequency range covered by the MWA
\item measure both XX and YY polarisations so Stokes parameters can be calculated
\item cover as many pointings as possible to fully explore predictive powers of any model
\item compare any measured beams to the FEE model.
\end{itemize}
For a single experiment, this is a daunting list. In this paper, we make a first step towards onsite beam measurements, and choose to limit ourselves to checking for variance in tile to tile beam patterns. The goals of this paper are: make an initial assessment of the accuracy of the new FEE model in describing the primary beam shape generated by onsite MWA tiles; quantitatively measure how similar primary beam shapes are from tile to tile; use the results to inform follow-up experiments covering the items on the list above.

The paper is organised as follows. In Section~\ref{sec:exsetup}, we describe our experimental setup and data collection method. In Section~\ref{sec:analysis} we detail our data analysis, and present our results, along with some experimental biases, in Section~\ref{sec:results}. We discuss our results and outline a possible future direction in Section~\ref{sec:discussion}.

\section{Experimental Setup and Data Collection}
\label{sec:exsetup}
Our experimental approach builds upon the work of~\citet{Neben2015}, however the basic premise of the experiment is worth repeating here. Any antenna beam shape can be mapped by measuring the power received when a transiting satellite passes, as long as one simultaneously observes using a reference antenna with a known beam response. The antenna under test (AUT), an MWA tile, and a reference antenna, receive powers $P_{\mathrm{AUT}}$ and $P_{\mathrm{ref}}$, respectively. The power received by each antenna is a combination of their beam responses, $B_{\mathrm{AUT}}$ and $B_{\mathrm{ref}}$, and the incident flux density from the satellite $F$, such that $P_{\mathrm{AUT}} = B_{\mathrm{AUT}}F$ and $P_{\mathrm{ref}} = B_{\mathrm{ref}}F$. The desired beam response can then be recovered through
\begin{equation}
B_{\mathrm{AUT}} = \dfrac{P_{\mathrm{AUT}}}{P_{\mathrm{ref}}} B_{\mathrm{ref}}.
\label{eq:beam}
\end{equation}
By observing enough satellite passes the beam response can then be measured across the entire sky. The key differences between our experiment and \citet{Neben2015} are:
\begin{itemize}
\item we conduct our measurements on four actual MWA tiles that form part of the telescope located at the MRO
\item our measurement system uses small, inexpensive, hand-held spectrum analysers and single-board computers to record raw power data from each tile
\item we use only open-source ephemeris and transmission-frequency data for the satellites in our analysis (as we did not have access to an `ORBCOMM user interface box'). 
\end{itemize}

\begin{figure*}[h!]
\centering
\includegraphics[width=1.95\columnwidth,clip,trim=130 50 80 50]{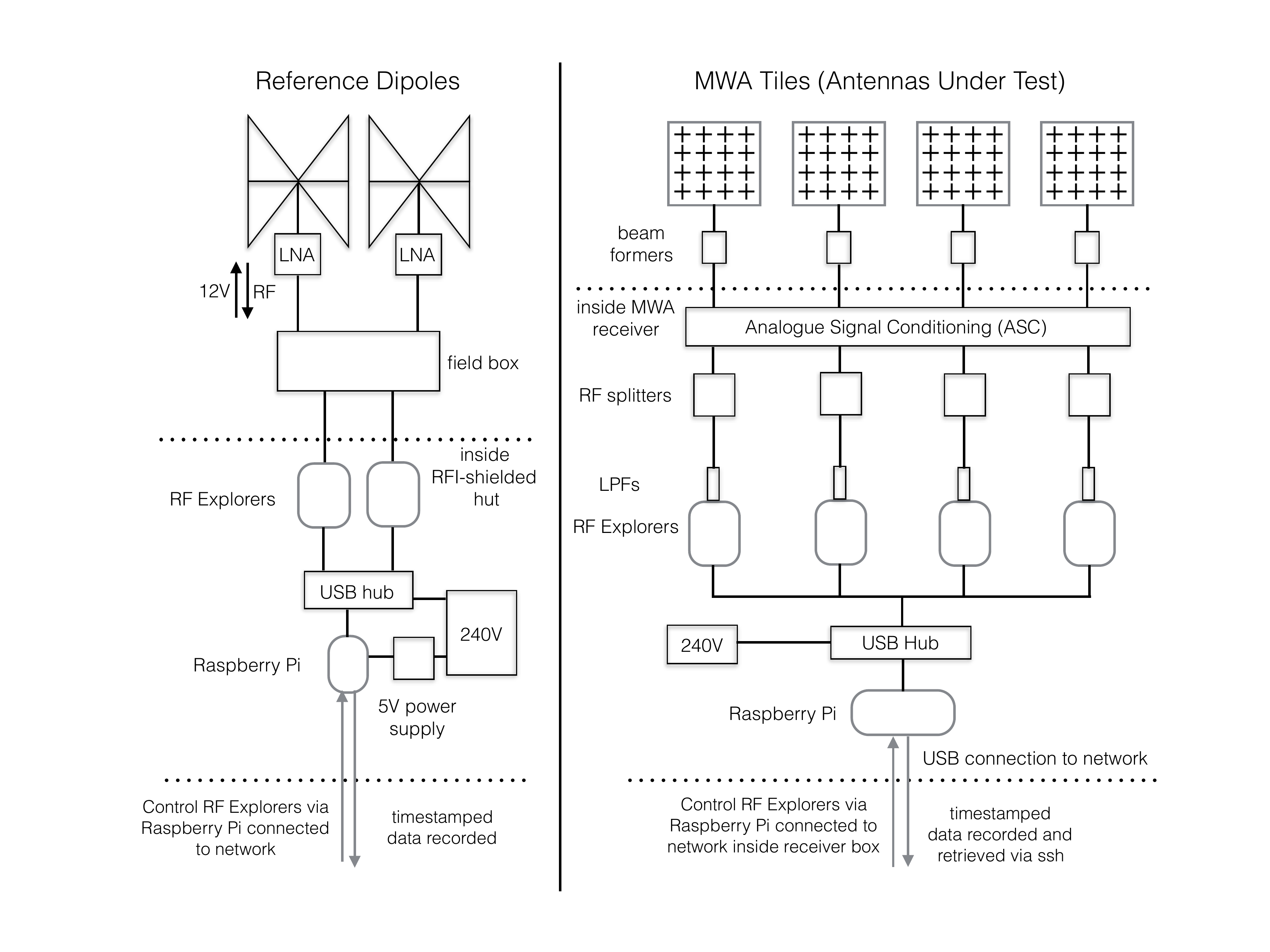}
\caption{\textsf{Block diagram of our experimental setup to measure four MWA tile beam shapes using ORBCOMM satellites. \textbf{Left:} The signal path for the reference dipoles. ORBCOMM signals are received by the reference dipoles where they are amplified by Low Noise Amplifiers (LNAs) attached to each dipole. Coaxial cables connected to bias-Ts in the field box both supply 12$\,$V DC to power the LNAs and also carry the Radio Frequency (RF) signal to the field box. Following further amplification by LNAs within the field box, the RF signal is fed via coaxial cable into the RFI-shielded hut where it forms the input to the RF Explorer spectrum analysers. The RF Explorers are connected via a powered USB hub to a single Raspberry Pi computer. The USB connection both powers the RF Explorers and provides the data transfer to the Raspberry Pi. The output of the RF Explorers (power measurements in dBm for 112 frequency channels at a rate of 7 samples per second) is recorded on the Rasperry Pi. The Raspberry Pi is accessed remotely via ssh from the MWA network to control recording and to transfer data for further processing. \textbf{Right:} The signal path for the Antennas Under Test (MWA tiles). For each MWA tile, the ORBCOMM signals are received by the dipoles and fed to the analogue beamformer. For our zenith-pointed observations all beamformer delays are set to zero. The signals are combined and transmitted via coaxial cable to an MWA receiver in the field. Inside the receiver the RF signal undergoes some amplification and filtering in the Analogue Signal Conditioning (ASC) stage before being split (resulting in a 3~dB drop in amplitude) and fed into a low-pass filter (LPF). From here the signal is fed to the RF Explorer which performs the spectral analysis and transmits the digital data via a powered USB hub to a Raspberry Pi computer. The Raspberry Pi is accessed remotely via ssh from the MWA network to control recording and to transfer data for further processing.}}
\label{fig:block_diagram}
\end{figure*}

A block diagram of our experimental setup is shown in Figure~\ref{fig:block_diagram}. We simultaneously record raw, radio-frequency power (RF) from two reference antennas and four MWA tiles. The reference antennas are required to remove the unknown intrinsic beam shapes and transmission powers of the individual ORBCOMM satellites. They each consist of a single MWA dipole sitting in the centre of a $5\,\mathrm{m}\times5,\mathrm{m}$ conducting mesh grid. We tap the RF from the tiles using beam-splitters inside the receiver box that houses the first stage of MWA signal conditioning (see \citealt{Tingay2013}). The RF is tapped after some initial analogue signal conditioning (including bandpass filtering and amplification) and prior to digitisation. The RF is low-pass filtered prior to input into the RF Explorer\footnote{\url{http://rfexplorer.com}} spectrum analysers, which record the raw RF data. The low-pass filtering is used to remove a known Radio Frequency Interference (RFI) signal generated by the RF Explorers, at around 400~MHz.

The tiles selected for testing were named S21-S24, which form part of the southern `hex' used in the compact configuration of the MWA (Wayth et al. in preparation). Tiles were removed from the observing program and pointed to zenith by setting all analogue beam-former delays to zero. This was done to maximise the number of satellite passes for one particular pointing, as we had limited time to conduct the experiment.

\begin{figure}
\centering
\includegraphics[width=1.0\columnwidth]{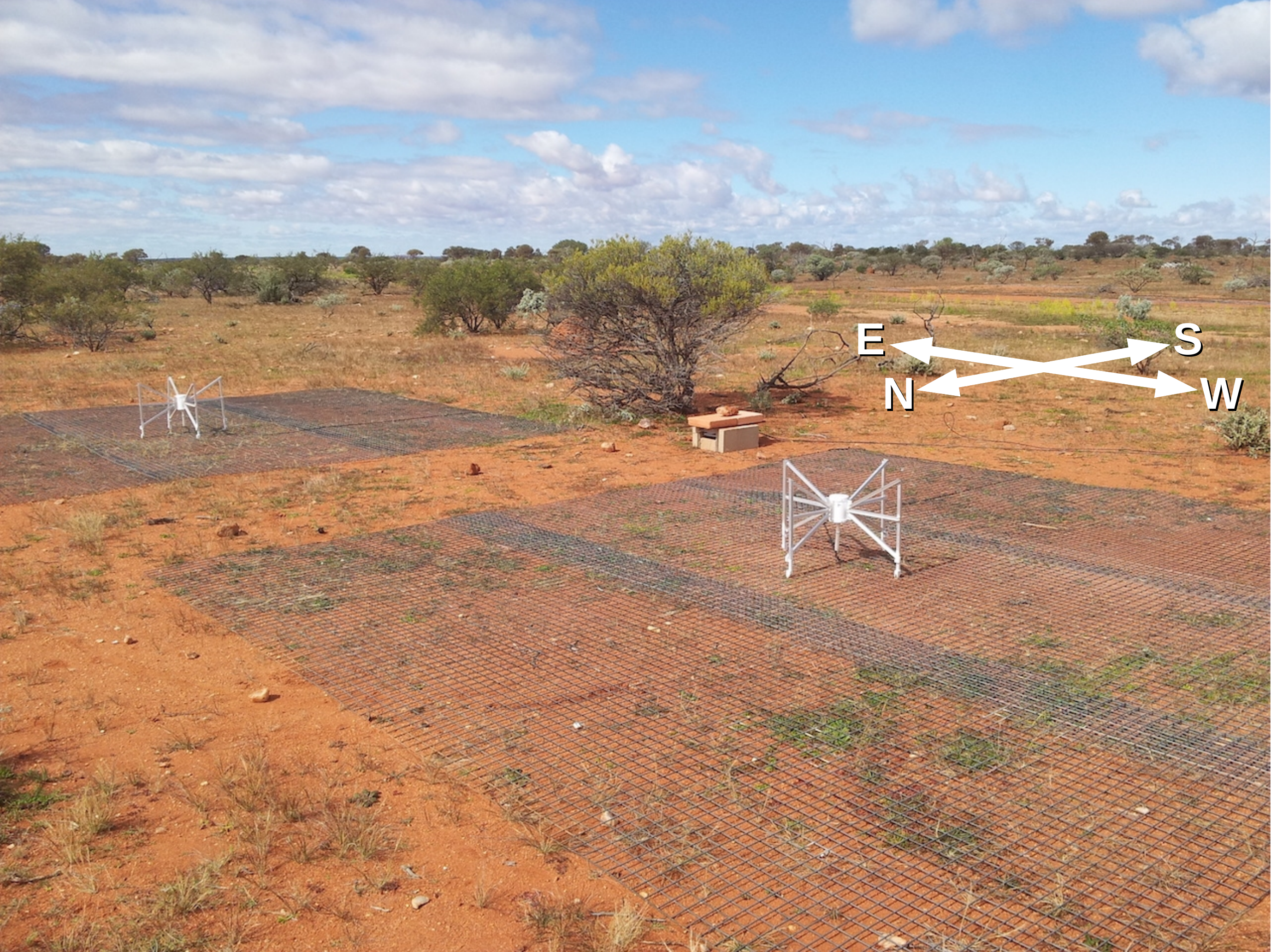}
\caption{\textsf{The reference antennas as deployed on site. The RFI-shielded box containing the power and data transfer electronics was protected by bricks, as can be seen near the image centre.}}
\label{fig:ref_antennas}
\end{figure}

Power and data transfer to and from the reference antennas was provided by a custom-built circuit, housed in an RFI-shielded box and placed in the field between the two antennas (see Figure~\ref{fig:ref_antennas}). This box was connected via coaxial cable to the RF Explorer and Raspberry Pi system set up inside the small RFI-shielded hut located approximately 50$\,$m away.

The RF explorers recorded 112 frequency channels between 137.150 and 138.550$\,$MHz at a spectral resolution of 12.5$\,$kHz. This automatically set the RF explorer to sample the power at temporal rate of 7$\,$Hz. We used Raspberry Pi\footnote{\url{https://www.raspberrypi.org}} (RP) single-board computers to record the data. We used the default Raspbian\footnote{\url{https://www.raspberrypi.org/downloads/}} operating system and a custom \texttt{python}\footnote{\url{http://www.python.org/}} script, using the \texttt{pySerial}\footnote{\url{https://pythonhosted.org/pyserial/pyserial.html}} module to control the RF Explorers. We utilised the \texttt{at}\footnote{\url{http://man.he.net/?topic=at&section=all}} command to schedule observing. We used the same NTP server to synchronise the two separate RPs. As the RPs were connected to the local network, which in turn was connected to the internet, were we able to remotely schedule observing once we left the site.

Using this setup we collected 403 hours of data between the 17$^{\mathrm{th}}$ of August and the 5$^{\mathrm{th}}$ of September 2017, while the MWA telescope was offline.

\section{Data Analysis and Method}
\label{sec:analysis}
\begin{figure*}
\centering
\includegraphics[width=1.95\columnwidth]{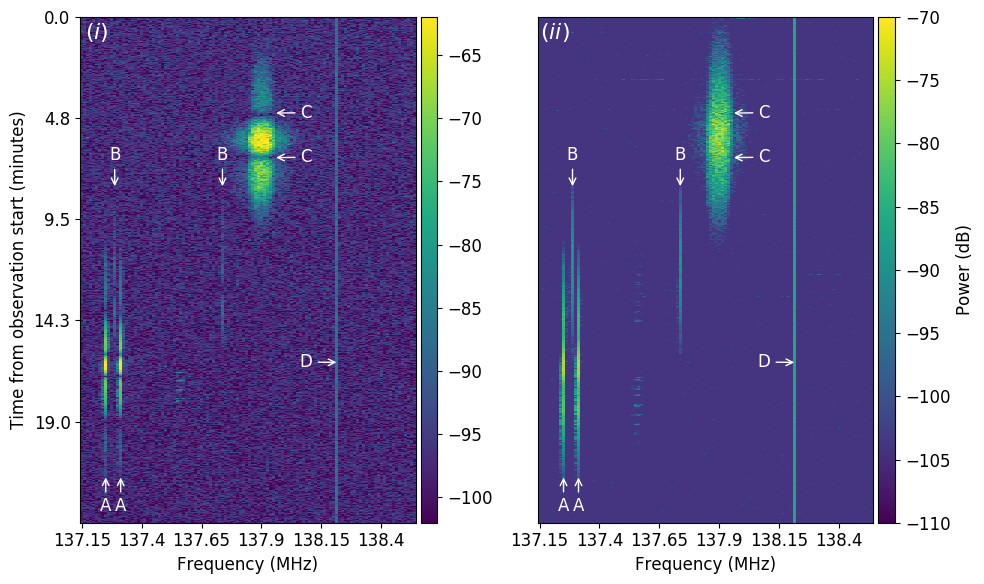}
\caption{\textsf{The raw data taken beginning 21:20 on 28/07/2017 for: (i) tile S24, (ii) reference antenna RF0. The colour scale on both plots spans the same dynamic range (40$\,$dB) to compare the signal to noise for each antenna. A number of features are annotated and labelled at the same position on each plot. ORBCOMM satellites typically emit in two distinct frequency channels; the A and B labels point out two different satellite passes. As each satellite rises and then sets (moves from top to bottom on the waterfall), the power seen by both S24 and RF0 increases and then decreases. The effects of the MWA primary beam can be clearly seen at C in (i). As the satellite passes through the nulls of the beam, the power seen drops to~$\sim$zero; this is not seen in (ii). This satellite pass is from a weather satellite with more broadband emission than an ORBCOMM satellite. These weather satellites were not used in our data analysis. Finally, D shows a frequency channel with constant power. We believe this is internally generated RFI from the RP itself.}}
\label{fig:waterfall}
\end{figure*}

An example of the raw data are shown in Figure~\ref{fig:waterfall}. Individual ORBCOMM satellite passes are clearly visible, and emit within a single channel due to the small transmission bandwidth. \citet{Neben2015} took advantage of an `ORBCOMM user inferface box', which is able to explicitly match the satellite transmission frequency to the correct satellite ephemeris. These boxes are not commercially available however. Instead, we pulled ephemeris data from the online service Space-Track\footnote{\url{https://www.space-track.org/auth/login}}. We then produced waterfall plots, such as those shown in Figure~\ref{fig:waterfall}, and matched the timing of the visible passes with satellites that were above the horizon. Although individual ORBCOMM satellites are known to periodically switch frequency channels, we found during our observations that most satellites tended to remain emitting at the same frequencies.

\subsection{Quality cuts}
\label{subsec:qualcuts}
We made a number of necessary quality cuts on the data. The first was to set a lower limit on the elevation of the satellite passes that we used. We set this to an elevation of $10^\circ$, as below this we found the results to have a very low signal to noise ratio.

The second cut was a noise cut on both $P_{\mathrm{AUT}}$ and $P_{\mathrm{ref}}$. As can be seen in Figure~\ref{fig:waterfall}, $P_{\mathrm{AUT}}$ and $P_{\mathrm{ref}}$ were well constrained to a single RF explorer frequency channel. To establish a noise floor, we took the data from the adjoining three frequency channels either side of the channel containing the pass, for the full duration of the pass, and took a median $\mu_{\textrm{noise}}$ and median absolute deviation $\sigma_{\textrm{noise}}$ of the data. We then made a criterion that $P_t > \mu_{\textrm{noise}} + \sigma_{\textrm{noise}}$, where $P_t$ was the power seen at any time-step within the pass. This procedure was completed for both the AUT and reference antenna; if either failed, the time-step data were discarded.

The third cut was a timing cut. As mentioned in Section~\ref{sec:exsetup}, we used the scheduling command \texttt{at} to run our controlling script. The reference antennas and AUTs were controlled by two separate RPs, that were set to the same local time via an internet sync on the Raspian OS. We assume that these two clocks were well synchronised, however each data dump from the RF explorers carried a time stamp, which revealed that at the start of each observation, a small timing offset of around 0.1-0.5$\,$s existed between the reference antenna and the AUT. This offset could be due to the \texttt{at} command itself, or perhaps because the RP controlling the AUT RF explorers had to launch multiple jobs simultaneously. A further complication arose in a faulty USB hub connecting the reference antenna RF explorers, which required us to restart the hub every half an hour.\footnote{The MRO is \textit{very} isolated, and it was quicker to spend the better part of a day finding a software solution than to drive and buy a replacement} As a result, each half hour observation had a differing timing offset between reference and AUT. We found that a robust way to treat the timing offset for each observation, reference antenna, and AUT, was to align the time series from the reference and AUT in such a way as to minimise the timing offsets $t_{\mathrm{off}}$. We then took a mean $\mu_{\textrm{t}}$ and standard deviation $\sigma_{\textrm{t}}$ of all $t_{\mathrm{off}}$, and required for a data point to match between the reference and AUT, that $t_{\mathrm{off}} < \mu_{\textrm{t}} + \sigma_{\textrm{t}}$. We found setting a single cutoff for $t_{\mathrm{off}}$ to be too inflexible given that the timing offsets varied, and often either under or over penalised data, giving poor results.

The final cut was a visual inspection of the results. As mentioned in Section~\ref{sec:analysis}, ORBCOMM satellites are known to periodically switch transmission frequencies. A number of passes $(\mathcal{O}(10)$) were clearly noise-like, which could be due to frequency channel misallocation. These passes were manually flagged and discarded.

After a combination of the number and duration of satellite passes $>10^\circ$ elevation during our observational window, and the time matching and noise cuts, 137 hours of data remained (34\% of our 403 observed hours).

\subsection{Map making}
\label{subsec:map_making}
To create beam maps, we used the satellite ephemeris to grid $P_{\mathrm{AUT}} / P_{\mathrm{ref}}$ for each time step onto a Healpix~\citep{Gorski2004} map with an $N$-side of 32. This corresponds to an angular resolution of 110 arcmins, which we found to give a good balance between detail on the sky, and the amount of signal integrated per pixel. Once gridding was complete, we were left with a distribution of values in each pixel that typically contained multiple outliers. We therefore took the median of each pixel, before multiplying by our model for $B_{\mathrm{ref}}$ to complete Equation~\ref{eq:beam}.

\section{Results}
\label{sec:results}
Whilst carrying out our analysis, we performed a null test using both reference antennas. As both reference antennas (hereto referred to as ref$_0$ and ref$_1$) should have identical beam responses, $P_{\mathrm{ref_0}} / P_{\mathrm{ref_1}} = 1$ should hold across all satellite passes. Upon investigation, we saw direction dependent deviations away from a ratio of one in the null test, beyond the error terms (as the raw RF data $P_{\mathrm{ref_0}}$, $P_{\mathrm{ref_1}}$ contain satellite beaming effects, we used the median absolute deviation of the distribution of measured values in each healpixel as an error term). This deviation away from unity prompted us to check the physical condition of the reference antennas. The onsite operational team took measurements and discovered that the ground meshes were not symmetric, and furthermore the dipoles themselves were not central within the ground screen. To mitigate these effects, we ran new reference antenna simulations.

\subsection{New reference models}
\label{subsec:reference_models}
\begin{figure*}
\centering
\includegraphics[width=1.6\columnwidth]{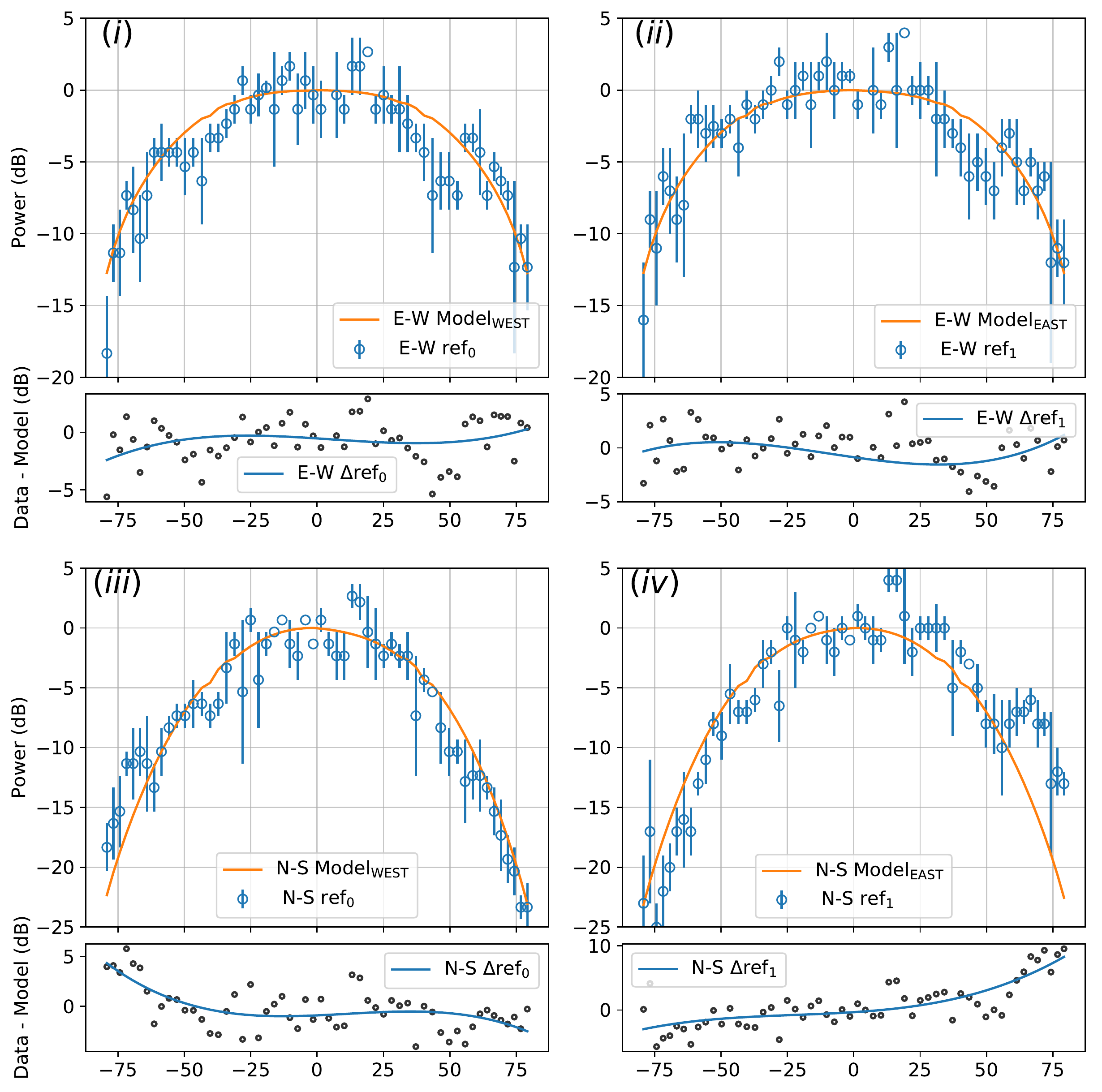}
\includegraphics[width=1.6\columnwidth]{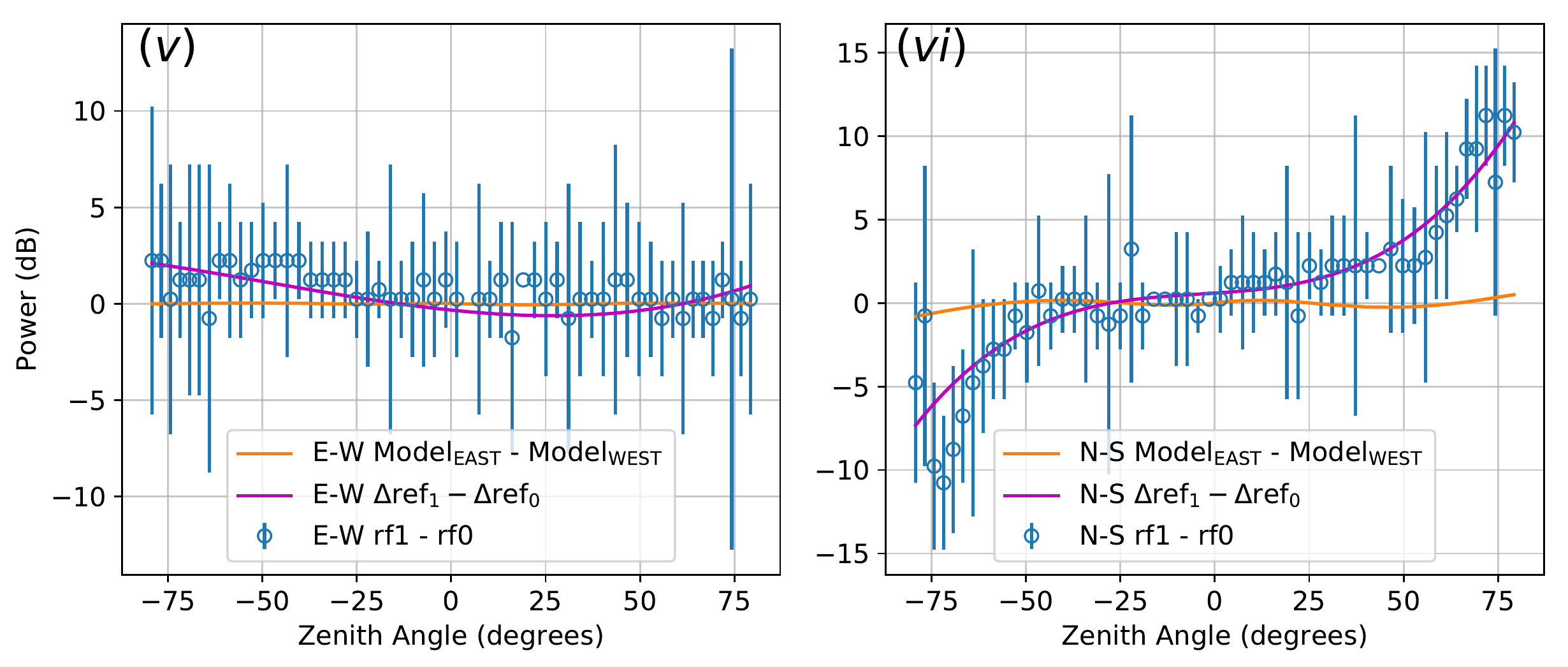}
\caption{\textsf{\textit{(i) \& (ii)}: in the upper panels, the raw RF data collected for ref$_0$ and ref$_1$, respectively, mapped onto a healpix grid and then sliced along the east-west direction (blue circles). Each data point is the median value in the healpixel, with median absolute deviation as the error bars. The EAST and WEST reference antenna models described in Section~\ref{subsec:reference_models} are over-plotted (orange line). In the lower panels, the difference between the RF data and the model are plotted (black circles), with a third order polynomial fit (blue line) over-plotted. \textit{(ii) \& (iv)}: same as (i) \& (ii), but in the north-south direction. \textit{(v) \& (vi)}: the null test in the east-west and north-south directions, respectively, performed by subtracting the ref$_0$ data from the ref$_1$ (blue circles). The errors bars come from a simple error propagation from plots (i)-(iv). The expected result of the null test from the FEKO models is also plotted (orange line), as well as the expected result given the fitted offsets found in (i)-(iv) (magenta line).}}
\label{fig:null}
\end{figure*}

We ran the same FEKO simulations used to generate the FEE, using the reference ground screen and dipole configurations as measured onsite, to generate new reference antenna models. Figure~\ref{fig:null} shows the raw RF data recorded (in subplots~\ref{fig:null}{\color{blue}(i)}-{\color{blue}(iv)}) for each reference antenna, after all quality cuts. We label the two FEKO reference antenna models EAST and WEST. The models were fit to the data by a single multiplicative gain factor through a least-squares minimisation, and show good agreement to the data, with the WEST model matching ref$_0$, and the EAST model matching ref$_1$. The plots show slices in east-west and north-south directions, taken by slicing healpix maps created as described in Section~\ref{subsec:map_making}. We gridded the FEKO models onto a healpix map for direct comparison. The difference between the model and the raw RF data is also plotted in subplots~\ref{fig:null}{\color{blue}(i)}-{\color{blue}(iv)}, with a third order polynomial fit. This functional form was chosen as it is smooth, and captures the behaviour of the offsets well. We label these fits to the offsets as $\Delta$ref$_0$ and $\Delta$ref$_1$. Subplots~\ref{fig:null}{\color{blue}(v)} and {\color{blue}(vi)} show the null test between the two reference antennas.

The test shows the models and data agree in the east-west slice, but a discrepancy is seen in the north-south direction. There are multiple factors that could cause the model to poorly describe the data, including: the ground could have a local gradient, pointing the antenna bore-sight away from zenith; the ground screen could be warped; the antennas could be faulty; the antennas could be coupled to their surrounding, or blocked in some way. Closer investigation of subplot~\ref{fig:null}{\color{blue}(iv)} reveals a bump in the data, away from the model, between a zenith angle of 50-75$^\circ$. Figure~\ref{fig:ref_antennas} shows that for the eastern tile, a rather large bush sits just to the south. It is quite possible that the bush is interfering with the both reference antennas, but more so for ref$_1$. Unfortunately, hardware limits such as cable length, and geographical limits of flat space close enough to the RFI-shielded hut providing power limited our deployment options.

To verify whether the deviations away from the models seen from the raw data are due to the reference antennas themselves, rather than the satellite beaming effects that are present within the raw RF data, the fits to the offsets taken in the lower panel were compared the null test. In~\ref{fig:null}{\color{blue}(v)} and {\color{blue}(vi)}, we plot $\Delta\mathrm{ref}_1 - \Delta\mathrm{ref}_0$, in the same way as the null test. Good agreement is shown between the model offset fits and the null test, giving us confidence that $\Delta$ref$_0$ and $\Delta$ref$_1$ can be used to predict systematic effects when generating beam maps, which we do in the following Sections.

\subsection{Tile maps}
\label{subsec:tilemaps}
\begin{figure*}
\centering
\includegraphics[width=1.6\columnwidth]{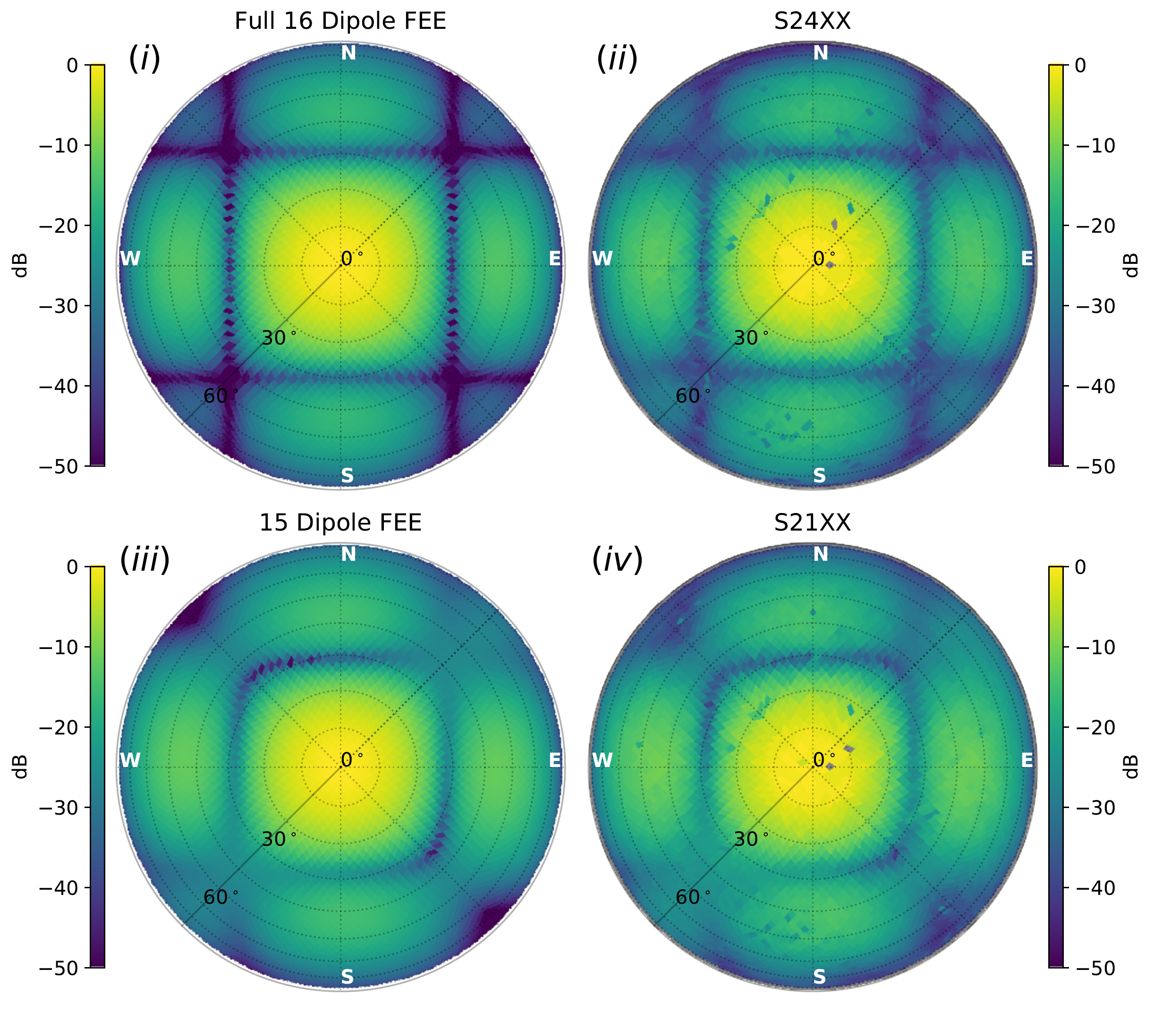} 
\caption{\textsf{\textit{(i)}: the zenith pointing FEE model for the XX polarisation with all 16 dipoles; \textit{(ii)}: the measured primary beam for tile S24; \textit{(iii)}: the zenith pointing FEE model for the XX polarisation with a central dipole missing; \textit{(iv)}: the measured primary beam for tile S21. All maps are normalised to zenith, and plotted with the same colour scale range for direct comparison.}}
\label{fig:S24}
\end{figure*}

We created maps using the method described in Section~\ref{subsec:map_making} using ref$_0$, as it showed smaller deviations away from the FEKO model than ref$_1$. An example map for tile S24 is compared to the FEE model in Figure~\ref{fig:S24} (top row). Both maps have been normalised to zenith. We normalised the measured map by fitting the map to the FEE model with a single gain factor using a least-squares minimisation, using a map cut-off of a zenith angle $<20^\circ$. This cut-off defines a region of the primary beam with high gain, as well as where the ORBCOMM signal transmission strength peaks. The primary beam shape is clearly captured in the map (see Figure~\ref{fig:S24}{\color{blue}(ii)}). Upon the creation of the beam map for S21, it was clear that the full FEE model was not a good match. As the MRO is situated in an inhospitable desert environment, components fail from time to time. It is standard practice at the MWA to regularly test components, and if the signal path to an individual dipole within a tile fails testing, it is excluded from the beamforming stage. With this in mind, we compared the map for tile S21 to FEE models where the gain of a single dipole was zero. We found that a model where one of the central dipoles was missing matched well, as shown in the bottom row of Figure~\ref{fig:S24}. The position of the missing dipole within the $4\times4$ arrangement greatly affects the resultant beam shape. After checking with the operations team, a dipole from the centre of S21 had indeed been excluded from the beamforming stage, matching our prediction.

\begin{figure*}
\centering
\includegraphics[width=1.4\columnwidth]{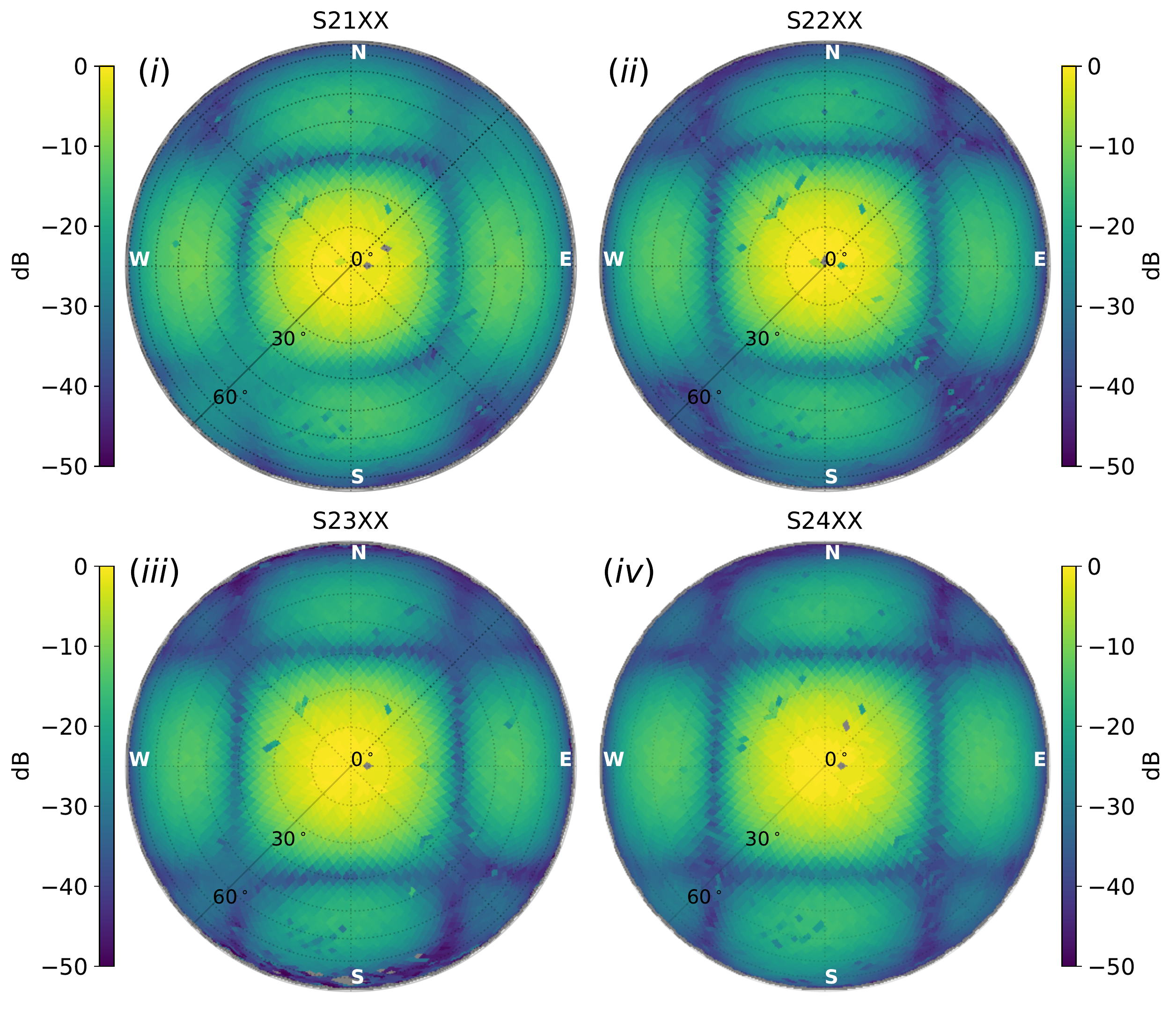}
\includegraphics[width=1.4\columnwidth]{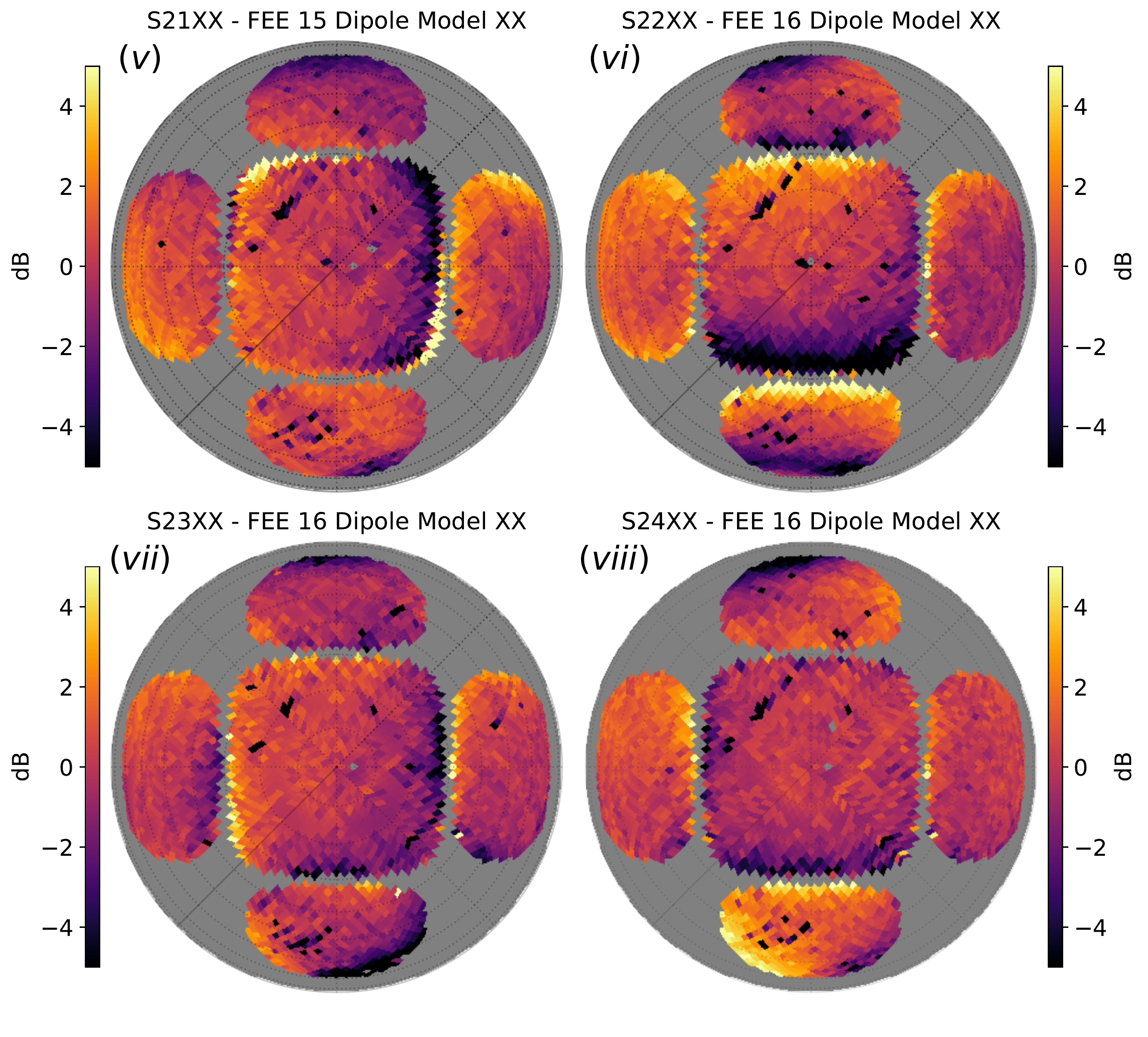}
\caption{\textsf{\textit{(i)-(iv)}: measured beam maps for tiles S21-S24, respectively; \textit{(v)-(viii)}: the ratio between the measured beam maps and the FEE. S21 is compared to the FEE model with a missing central dipole; all other maps are compared to the full 16 dipole FEE model. For clarity, the ratio maps are masked where the full 16 dipole FEE model beam is predicted to have $P < -30\,$dB.}}
\label{fig:good_maps}
\end{figure*}

Figure~\ref{fig:good_maps} compares the maps generated for tiles S21 - S24 in subplots ~\ref{fig:good_maps}{\color{blue}(i)}-{\color{blue}(iv)}, as well as the ratios\footnote{these are differences in units of dB, which are on a log scale, and so are in fact ratios} of these normalised maps to the FEE model, shown in~\ref{fig:good_maps}{\color{blue}(v)}-{\color{blue}(viii)}. S21 was compared to the 15 dipole FEE model, with the other tiles compared to the full 16 dipole FEE model. The ratio maps show that there is tile to tile variation. Interestingly, the ratio map for S22 suggests a north-south gradient across the central lobe. We would like to emphasize that as all maps were made using ref$_0$, any systematic errors from the reference antenna are identical within the maps, and so the variations seen are purely from the tiles and their signal paths.

\begin{figure*}
\centering
\includegraphics[width=1.4\columnwidth]{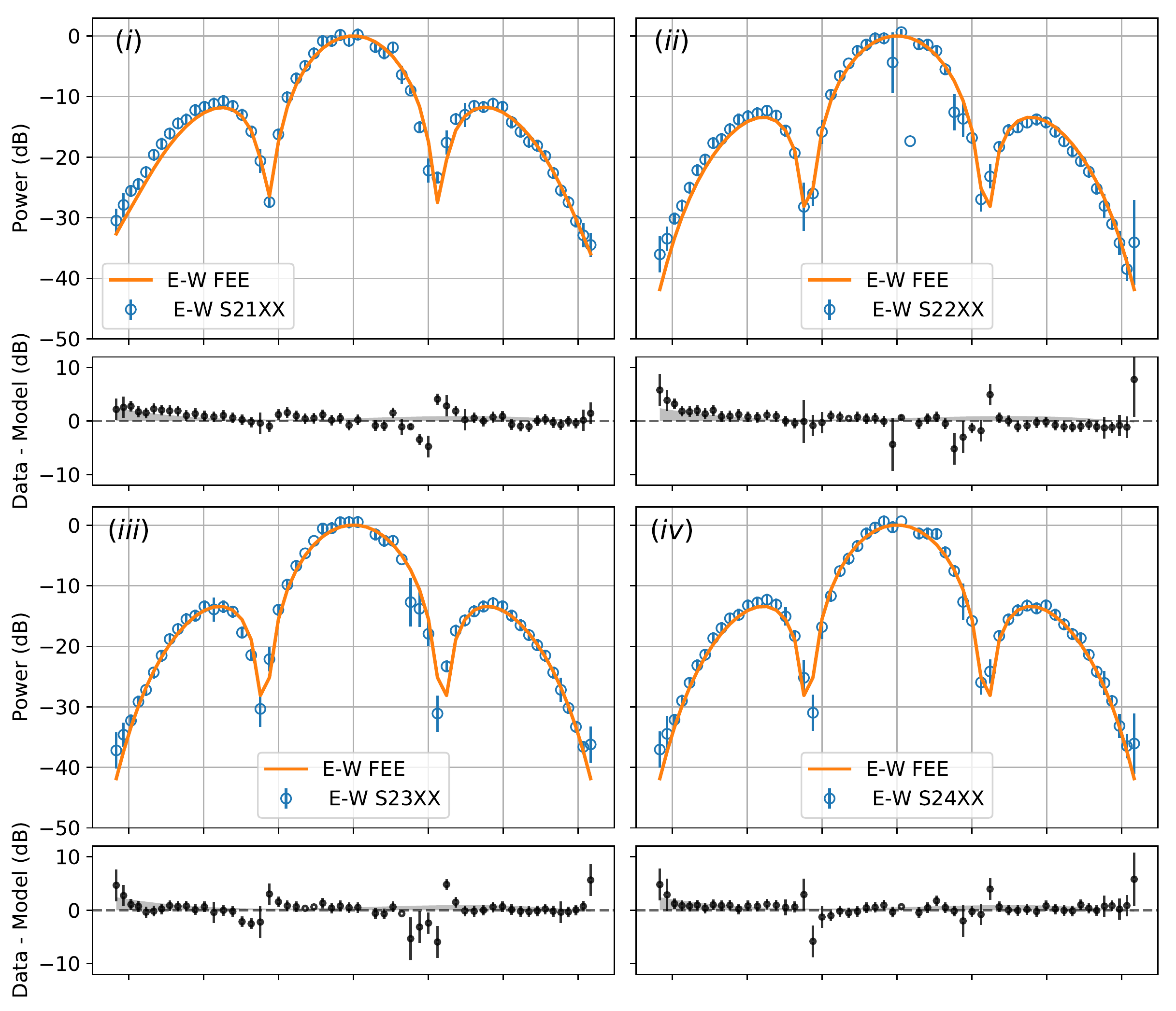}
\includegraphics[width=1.4\columnwidth]{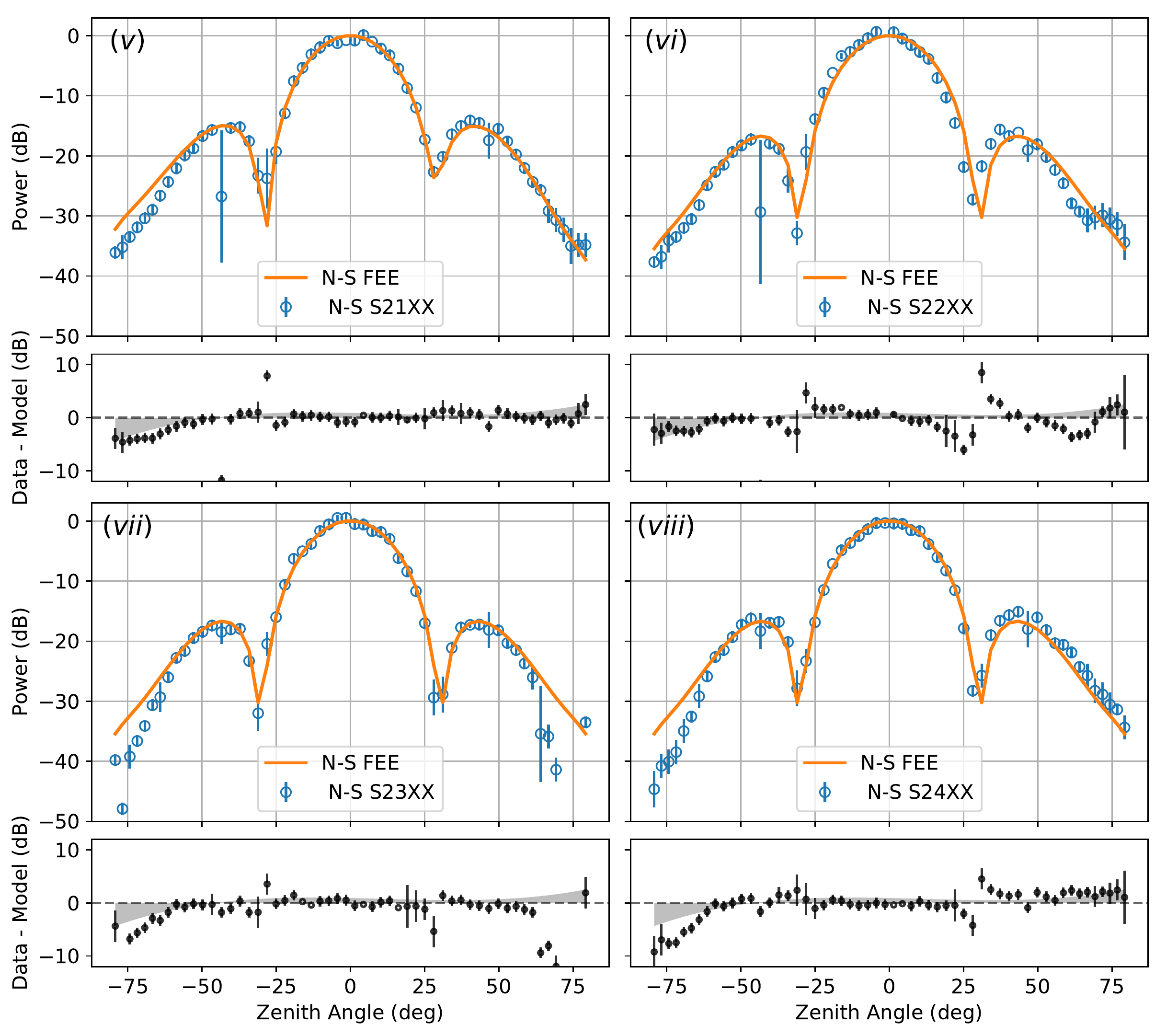}
\caption{\textsf{East-west \textit{(i)-(iv)} and north-south \textit{(v)-(viii)} beam map slices for tiles S21-S24, respectively. For each subplot, the upper panel shows the healpixel median value and absolute deviation (blue circles with error bars), with the FEE model over-plotted (orange line). In the lower panel, the difference in dB between the FEE model and the data are plotted (black circles), with a reference dashed line at zero (a ratio of one). Again, S21 is compared to the FEE model with a missing central dipole; all other maps are compared to the full 16 dipole FEE model.The grey shaded areas are an estimate of the systematic error introduced by ref$_0$, using the fit to the offsets from the FEKO reference antenna model found in Figure~\ref{fig:null} (explicitly, we have used $1 / \Delta\mathrm{ref}_0$).}}
\label{fig:good_slices}
\end{figure*}

In Figure~\ref{fig:good_slices}, we compare the FEE model to the beam maps for S21-S24 in the east-west direction (\ref{fig:good_slices}{\color{blue}(i)}-{\color{blue}(iv)}) and the north-south direction (\ref{fig:good_slices}{\color{blue}(v)}-{\color{blue}(viii)}). Excellent agreement to the FEE model (15 dipole model for S21 and 16 dipole model for all other tiles) is shown for all tiles in the east-west direction, even down to $-30\,$dB and below. As predicted from the null test in Section~\ref{subsec:reference_models}, the maps perform worse in the north-south direction. The model and data still agree well in the central lobe (with exception of S22 which shows a gradient away from zero), but disagree out in the sidelobes. For each comparison, we have also plotted the difference in dB between the measured map and the FEE model. Plotted with the differences are shaded grey areas. These are the probable biases introduced by ref$_0$, estimated using the fitted deviations away from the WEST reference antenna model, $\Delta\mathrm{ref}_0$.

\section{Discussion and next steps}
\label{sec:discussion}
The first step in furthering this experiment would be to move the reference antennas, replace the ground mesh and make sure the layout of the reference antennas are identical. The next obvious step is to also test the YY polarisation: this would allow us to compare to polarisation leakage tests. Increasing the number of tiles under test would allow an instrument-wide characterisation of the tile-to-tile variation.

The method we have developed is simple to implement, but as noted in Figure~\ref{fig:block_diagram}, we currently split the RF signal to measure it, which results in a $3\,$dB drop in signal through to the correlator. Further work is required to assess the impact this has on the noise floor of the instrument and in turn on the scientific outcomes of affected programs. If the scientific impact is acceptable, we plan to collect ORBCOMM data during normal operations. Over the course of an observing season, a greater number of pointings can be explored. Leakage seen from calibrated visibilities can then be compared to that inferred from the beam measurements, as a function of pointing direction. We could also measure the stability of the beam shapes over time. Not only could this help improve beam models, but it could potentially inform calibration pipelines.

During the analysis of the data, we found that our original reference beam models did not match our measured reference beam shapes. We observed that the exact component sizes, layout, and surrounding environment all contributed to changing the in situ beam shape. This result emphasizes that antenna configurations need to be accurately deployed to obtain the expected results, and the need to conduct onsite measurements to be sure that the behaviour in the field matches laboratory measurements.

Considering all 4 maps together, and all pixels down to $10^\circ$ elevation ($4\times5056$ pixels), we find the median absolute offset from the FEE model to be $1.4\pm1.1\,$dB, where the error is the median absolute deviation. If we only consider the east-west slices (shown in Figure~\ref{fig:good_slices}{\color{blue}(i)}-{\color{blue}(iv)}), we find the median absolute offset from the model to be $0.7\pm0.4\,$dB. We therefore conclude that the MWA analogue beamformers reliably create the same zenith beam shape to within~$\sim 1$dB, and that the FEE model is able to predict this beam shape even with a missing dipole. As mentioned in Section~\ref{sec:intro}, the major advantage of having the same beam shape for each receiving element is the reduction in the complexity of the software required to reduce the collected visibilities. Not only does this reduce computational loads, but the simpler the calibration and imaging approach, the easier it is to understand the impact on the created data products. This directly relates to the upcoming SKA science data processor\footnote{https://www.skatelescope.org/sdp/}, which will deliver calibrated science-ready data products in `soft-realtime'. Any real time processing instrument benefits from a calibration scheme and hardware build that have a complimentary philosophical design.

It is still unclear, however, if the beam-to-beam variations measured here are small enough to not affect a possible EoR detection. As discussed in Section~\ref{sec:intro}, there are a number of ways beam errors can affect the data. We plan on using the beam variations seen here in future simulation work to assess the possible impacts on EoR science: in particular, the impact of missing dipoles from tiles.

\section{Conclusions}

We have measured the XX polarisation primary beam shape for 4 onsite MWA tiles. The coverage of the ORBCOMM constellation above the MRO allows full sky coverage down to an elevation of $10^\circ$, if one can continually observe with a single pointing over a period of two weeks. The maps have a dynamic range of $>30\,$dB, and show good agreement with the cutting-edge FEE model of \citet{Sokolowski2017} in the east-west direction. Geographical and hardware limitations reduce that agreement, and the reliability of the maps, in the north-south direction.

We observe that the FEE model is able to accurately describe the zenith beam pointing of an MWA tile, even when one dipole is missing. We also demonstrate the strong dependence of the primary beam shape on the number of contributing dipoles. This is important since the MRO is an inhospitable environment and it is common for dipoles to fail.

We have developed a method for collecting raw RF data from the tiles using relatively inexpensive and commercially available RF Explorers and Raspberry Pis. Most importantly, not only can all of this equipment easily fit within a receiver box, it can be remotely controlled and scheduled. This will allow us to easily expand this method to measure both instrumental polarisations and multiple beam pointings in future work, enabling a better understanding of instrumental performance, which will benefit all scientific programs of the MWA.

\section{Acknowledgements}
\label{sec:acknowledgements}
J.~L.~B.~Line would like to thank the MWA operations team; without their engineering expertise and tireless efforts this work would never have been completed. This scientific work makes use of the Murchison Radio-astronomy Observatory, operated by CSIRO. We acknowledge the Wajarri Yamatji people as the traditional owners of the Observatory site. Parts of this research were conducted by the Australian Research Council Centre of Excellence for All-sky Astrophysics (CAASTRO), through project number CE110001020.

\bibliographystyle{pasa}
\bibliography{Mendeley}

\end{document}